\documentclass[runningheads]{llncs}
\usepackage[T1]{fontenc}
\usepackage{graphicx}
\begin{document}
\title{Investigation of Energy-efficient AI Model Architectures and Compression Techniques for "Green" Fetal Brain Segmentation}
\titlerunning{"Green" Fetal Brain Segmentation}
%
\author{
Szymon Mazurek\inst{1,2}\orcidID{0009-0006-7557-0157} \and
Monika Pytlarz\inst{1}\orcidID{0000-0001-5319-1769} \and
Sylwia Malec\inst{1}\orcidID{0000-0002-3603-9930} \and
Alessandro Crimi\inst{1,2}\orcidID{0000-0001-5397-6363}}
\authorrunning{S. Mazurek et al.}
%
\institute{
1. Sano Centre for Computational Personalized Medicine, Nawojki 11, 30-072 Cracow, Poland 
\url{https://www.sano.science} \\
2. AGH University of Science and Technology,
Adam Mickiewicz Avenue 30,  30-059 Cracow, Poland \url{https://www.agh.edu.pl}\\
\email{a.crimi@sanoscience.org}}
\maketitle              
\begin{abstract}
Artificial intelligence have contributed to advancements across various industries.  However, the rapid growth of artificial intelligence technologies also raises concerns about their environmental impact, due to associated carbon footprints to train computational models. 
Fetal brain segmentation in medical imaging is challenging due to the small size of the fetal brain and the limited image quality of fast 2D sequences. Deep neural networks are a promising method to overcome this challenge. In this context, the construction of larger models requires extensive data and computing power, leading to high energy consumption. Our study aims to explore model architectures and compression techniques that promote energy efficiency by optimizing the trade-off between accuracy and energy consumption through various strategies such as lightweight network design, architecture search, and optimized distributed training tools. We have identified several effective strategies including optimization of data loading, modern optimizers, distributed training strategy implementation, and reduced floating point operations precision usage with light model architectures while tuning parameters according to available computer resources. Our findings demonstrate that these methods lead to satisfactory model performance with low energy consumption during deep neural network training for medical image segmentation.

\end{abstract}

\section{Introduction}
\subsection{Fetal Brain Segmentation}
Magnetic resonance imaging (MRI) is a popular non-invasive method for evaluating the development of the central nervous system of the fetus during pregnancy. In recent years, neuroimaging has become popular for studying the fetal brain. However, manual segmentation of the structures of the fetal brain is time-consuming and subject to variability between observers. Therefore, researchers have used artificial intelligence  to automate and standardize this process \cite{Ciceri2023}.
Fetal brain segmentation faces challenges due to the small size of the fetal brain, extra tissues that need to be distinguished, motion artifacts, and limited image quality from fast 2D sequences. 
Additionally, only narrow datasets that vary in image acquisition parameters are publicly available, which complicates the training of the Deep Learning (DL) algorithm. In terms of DL methods, various approaches have been tested, including unsupervised training, atlas fusion, deformation, and parametric methods. Many methods incorporate preliminary steps, such as fetal brain location and region of interest (ROI) cropping. Super-resolution reconstruction algorithms have enabled 3D segmentation, but standardization of these techniques is lacking. The majority of techniques make use of convolutional neural networks (CNNs), especially the U-Net architecture. Reconstruction algorithms and transfer learning methods are widely used to address data issues. Due to their ability to minimize partial volume effects, 3D segmentation techniques are becoming more and more common; but their performance depends on image quality \cite{Ciceri2023}.

\subsection{"Green" Deep Learning}
Over the past decade, significant advances have been made in artificial intelligence   and machine learning (ML), driven largely by the accessibility of large datasets and the rise of DL. Deep neural networks (DNNs) play a key role in DL, and a prominent focus while developing a new DNN architecture is on achieving state-of-the-art (SOTA) results. However, the progress of the current SOTA is often achieved by increasing the complexity of the model, leading to a 300,000-fold increase in computational load in a span of six years \cite{schwartz}. Although these networks deliver impressive results, improving their capabilities requires substantial data and computational resources, resulting in a high use of energy with notable economic and environmental implications. Consequently, there is a critical need for energy-efficient DL to address concerns related to finances, ecology, and practical usability \cite{mehlin2023towards}. Moreover, these models are extremely data-hungry, making its direct application to tasks, such as fetal brain segmentation, difficult. As a result, the exploration of their lightweight counterparts becomes a reasonable step towards addressing climate change; they will also facilitate the application of SOTA medical image analysis techniques in scenarios where energy is limited or costly, such as in developing nations and on portable devices relying on batteries. Moreover, they will encourage the adoption of improved training methods to replace the current conventional approach, which typically involves extensive searching for optimal hyperparameters and a trial-and-error process.\\
The energy efficiency of deep learning could be classified into five categories: robust information technology  infrastructure, modeling, efficient data use, energy-saving training, and inference. \\
The foundation for effective running of DL models is a reliable software and hardware system, including a careful selection of libraries and the use of graphics processing units, tensor processing units, or neuromorphic computing \cite{mehlin2023towards,menghani2021efficient}. 
Another crucial component in minimizing computational effort is the selection of the right architecture. Modeling efficiency can be achieved through the use of compact neural networks and various automation and assembly techniques \cite{menghani2021efficient,xu2021survey}. 
Even lightweight and optimized architectures often require a large amount of data to achieve peak performance. To reduce the data-related cost of training, data augmentation and active learning can be used \cite{menghani2021efficient,xu2021survey}. 
Various techniques have been suggested to reduce the expense of training itself, such as different initialization techniques, normalization, progressive training, and mixed precision training \cite{menghani2021efficient,xu2021survey}. 
After successful training, the energy efficiency of inference should be considered. Common compression methods contain quantization, pruning, low-rank factorization, deployment sharing, and knowledge distillation \cite{menghani2021efficient,xu2021survey}.

\subsection{Contribution}We aim to evaluate the environmental impact of deep learning by looking at energy consumption during model training. Existing solutions concentrate only on one category, such as architecture design \cite{wu2023lightnet} or hardware acceleration \cite{xiong}. In contrast, our study intends to thoroughly examine the impact of different optimization techniques, starting from architecture selection, efficient data usage, and training acceleration, leading to evaluation of the model's environmental effect. We investigated a variety of energy-efficient techniques to apply to the segmentation of the fetal brain in MRI, with the goal of developing a model with the best ratio of Dice score to energy consumption. 

\section{Methods}
\subsection{Experimental design}
At first, we chose U-Net \cite{unet} as our baseline due to its ubiquitous usage in medical image segmentation tasks. After applying pre-processing techniques, we evaluated various models and selected the one that obtained the best ratio of Dice score to consumed energy during training. The impact of energy-efficient methods was evaluated only on the selected model.

\subsection{Experimental setup and hardware}
All experiments were carried out using Python 3.11.5 with Pytorch 2.0.1 \cite{PaszkeTorch} and Lightning 2.0.3 libraries \cite{falcon2019lightning} with CUDA 11.7. Data loading and processing was performed with the Monai 1.2.0 \cite{CarsodoMonai} library. Energy usage was tracked using Codecarbon 2.3.1, and logged with Weights and Biases 0.15.4. We use an HPC environment containing 4 Nvidia A100 GPUs with 40GB of RAM memory, 120 cores of 2 AMD EPYC 7742 64-core processor and up to 500 GB of RAM memory.

\subsection{Data Source and Pre-processing Techniques}
For our project, we incorporated the Openneuro dataset, a library of 1241 manually traced fetal fMRI images of 207 fetuses \cite{Rutherford_2021}.
To comply with BIDS standards, the authors merged 3D volumes (raw and mask) into a 4D time series file \cite{Turk_2019}. The masks were drawn in a single volume from a period of fetal stillness. Within pre-processing, we cleaned the data from files missing matching masks .nii or raw .nii, extracted 3D volumes from functional MRI (fMRI) times series, and sliced 3D volumes into 2D pairs of slices raw vs. mask. 
Furthermore, we used a second dataset consisting of fMRI, T2-weighted, and diffusion-weighted MRI scans \cite{challenge3}. Then we merged these datasets, doing the subject-wise division. 
In total, we obtained 40945 2D slices, which were used for the experiments. Data were split by patient into training, validation, and test subsets. For the test 10\% patients were randomly chosen. Another 10\% was allocated from the remaining patients for validation. The rest was used during training.

\subsubsection{Data caching and loader configuration}
During neural network training, the speed with which the data can be provided to the model is often a bottleneck. Training on large datasets involves a lot of I/O operations, as the data needs to be loaded from the memory. Also, the pre-processing applied to the data on-the-fly slows down the process. This can be alleviated by using data caching in memory, especially if computational resources allow it. We evaluated the potential solution to this problem, Monai's CacheDataset abstraction. It first preloads the data into RAM and applies deterministic transformations, resulting in gains in model throughput at the cost of increased memory consumption.

Furthermore, we examined the impact of data-loading abstractions offered by popular machine learning frameworks such as Pytorch:
\begin{itemize}
    \item Number of workers determines the number of concurrent processes involved in accessing the data.
    \item Data prefetching is an operation of loading and pre-processing the data by each worker into a buffer, from where the data are immediately accessed when the model requests for it.
    \item Persistent workers option is an approach to handle worker processes. With this option selected, worker processes are not destroyed upon completion of an epoch (iteration thought entire dataset). This allows to reduce the time needed to spawn those processes.
    \item Memory pinning enables the allocation of a predetermined memory subspace from which the transfer of data to GPU is increased.
\end{itemize}

\subsubsection{Hyperparameter search techniques}
We adopted the automatic learning rate tuning offered by the Lightning API to select the initial learning rate \cite{falcon2019lightning}. This was the only chosen automatic tuning method, as the full architecture and hyperparameter search are compute-costly procedures.

\subsubsection{Data Augmentation}
Medical images can take advantage of number data augmentation techniques for natural image analysis such as geometric transformations (rotations, horizontal reflections, cropping, shifting), the addition of random noise, or gamma correction \cite{hussain2017differential}. Medical image datasets should not have been augmented via transformations influencing the color such as the modification of the saturation or the hue of the natural image; therefore, we applied simple geometric transforms, rotations, and flips, to reduce overfitting.

\subsection{Model Architectures and Selection}
We have conducted a review of lightweight network architectures analyzing them within four groups following Andrii Polukhin's division for tinyML \cite{youtube}: 
\begin{itemize}
\item MobileNet Family (models based on MobileNet v1 (2017)), 
\item Model Scaling Formula Family (models based on EficientNet v1 (2019)), 
\item Group Convolution Family (models based on CondenseNet (2017) with the MicroNet (2021) as the newest variation), 
\item Squeeze n Excitation Family (models based on SqueezeNet (2016)).
\end{itemize}
We searched whether these networks have been implemented both for classification and segmentation tasks. We have compared the number of parameters, the performance reported on the benchmark datasets, and the original DL library used to write the implementation. We have selected the following architectures to compare with UNet (30M params, PhC-C2DH-U373 cell dataset, intersection over union score - IoU - 0.9203) taken as a baseline model: 
\begin{itemize}
\item MobileNetV3-small (2.94M params, Cityscapes dataset 0.6838 IoU) \cite{howard2019searching},
\item MicroNet (5.36M params, Austin City image dataset, 0.7559 IoU) \cite{chen2018miniaturized},
\item EfficientNet (6.63M params, Flowers dataset, 0.4920 IoU) \cite{tan2019efficientnet},
\item Squeeze-UNet (\textbf{2.59M} params, CamVid dataset,  IoU 0.6160) \cite{beheshti2020squeeze}, 
\item Attention-Squeeze-UNet (2.6M params, PH2 dataset, \textbf{IoU 0.8753}) \cite{pennisi2022skin}.
\end{itemize}
In addition to the listed architectures, we adopted the project "Efficient-Segmentation-Networks Pytorch Implementation" with several model implementations to test them during our experiments: SQNet, LinkNet, SegNet, ENet, ERFNet, EDANet, ESPNetv2, FSSNet, ESNet, CGNet, DABNet, ContextNet and FPENet \cite{Efficient-Segmentation-Networks}.

\subsection{Model Optimization Techniques}
\subsubsection{Quantization for Training and Inference}
Quantization is a technique for reducing model size that converts model weights from high-precision floating point to low-precision floating point or integer representations, such as 16-bit or 8-bit. By converting the weights of a model from a high-precision to a lower-precision representation, the model size and inference speed can be increased without sacrificing too much precision. Additionally, quantization improves the efficacy of a model by reducing memory bandwidth requirements and increasing cache utilization \cite{cai2017deep}. However, quantization can introduce new challenges and trade-offs between accuracy and model size, especially when using low-precision integer formats such as INT8.

\subsubsection{Loss Functions}
The choice of the loss function can help the model both to achieve better final performance and to increase convergence speed. In the experiments, we evaluated the most popular loss functions used in training neural networks for segmentation, such as Dice, Binary Cross Entropy (BCE), and Matthews Correlation Coefficient (MCC).

\subsubsection{Pruning}
Pruning is a known technique to reduce the size of the model by removing a subset of parameters. This is intended to increase throughput and lower the computation requirements. Usually, this reduction comes with the cost of reduced performance; therefore, the procedure usually involves iterations of applying pruning followed by fine-tuning the model to regain the performance. 
Pruning can be classified as unstructured or structured. Unstructured pruning aims to remove chosen weights without altering the network structure.
The objective behind the use of structured pruning is to physically remove a group of parameters, thus reducing the size of neural networks. It also involves establishing the importance of the parameters to prune and the relationships between them. This is achieved using various algorithms, including the method evaluated in this study, a Dependency Graph (DepGraph). This method explicitly models the dependency between layers and comprehensively group coupled parameters for pruning, showing great results on benchmark tasks \cite{fang2023depgraph}.

\subsubsection{Gradient Averaging}
Gradient averaging stabilizes and speeds up training in distributed settings, when training CNNs with batch gradient descent. The process involves two main steps: forward and backward pass. During the forward pass, the model processes a batch of data and calculates an error compared to the desired results. In the backward pass, gradients with respect to all model parameters are computed and used to update the model weights iteratively until it converges to a local minimum in the error function.
From the available methods, stochastic weight averaging (SWA) was chosen for evaluation \cite{IzmailovSWA}. This method tracks the learned parameters for every epoch as the training nears the end and replaces the final ones with the average of them. Research has consistently shown that performance gains were observed in various problems solved using neural networks with nearly no additional computational costs.

\subsubsection{Choosing optimizer and its configuration}
To perform optimization in neural networks, various algorithms were proposed. The choice of the optimizer and its hyperparameters has a significant impact on the convergence of the model, training time, and generalizability.\\
We oriented ourselves towards adaptive optimization algorithms, which are able to tune the learning rates per parameter during training based on gradients. They are less influenced by the initial choice of learning rate, as it is only used mostly as an upper limit value for the aforementioned choice of the one specific for given parameter.
Therefore, we sought to examine the Novograd optimizer, a relatively novel method for adaptive optimization\cite{GinsburgNovograd}. 
This algorithm allows for adaptive parameter update and reduces
the memory footprint of the Adam optimizer by half. It was also shown to be more
robust to the choice of learning rates, therefore being the go-to choice when the
data are unknown or when the cost of hyperparameter search is to be avoided. It can also be paired with AMSGrad \cite{PhuongAMSGrad}, a solution to problems with convergence of Adam optimizer where learning rate updates were too aggressive, the closer the optima.
We also evaluated the influence of the exponential decay rate for the first and second moment estimates ($\beta_1$ and $\beta_2$).

\subsubsection{Exploring PowerSGD}
Vogels, Karimiredd and Jaggi (2019) proposed a promising optimization technique: PowerSDG is a low-rank gradient compressor based on power iteration that compresses gradients quickly, aggregates them using all-reduce, and achieves test performance comparable to stochastic gradient descent (SDG) \cite{vogels2019powersgd}.

\subsubsection{Distributed data parallel configuration}
Distributed data parallel (DDP) is a method of conducting distributed training by copying the model to each compute device, performing the forward pass on subset of training batches, and accumulating the results in the main process to update the model parameters. This strategy allows for configuration that potentially leads to training speedups and efficiency improvements. 
The static computational graph in neural network training is suitable for situations where the input shape and number of parameters in the model do not change, as function tracing is done only once, resulting in reducing the time necessary for every iteration of training. We also reduced the precision of the operations to 16-bit, lowering the memory requirements to increase the model throughput.

\subsubsection{Distributed Strategy - Comparative Analysis of Training Speedup with Bagua}
Bagua is a set of distributed training algorithms developed by Kuaishou Technology and DS3 Lab from ETH Zurich \cite{GanBagua}. The authors have demonstrated that the usage of these algorithms benefits the training speed compared to other strategies, including Pytorch DDP. The solutions offered were incorporated via the lightning-bagua library, a plugin that allows us to use these algorithms with the Pytorch Lightning framework \cite{falcon2019lightning}. 
The algorithms offered by Bagua are related to distributed communication (bytegrad, asynchronous model average, improved all-reduce, low-precision decentralized SGD) and optimization algorithm (QAdam \cite{TangQAdam}).

\subsection{Energy Monitoring and Analysis}
For tracking the relationship between performance and energy consumption, we measured the total energy consumed by the hardware used in training (CPUs, GPUs, and RAM memory) during the evaluation process. We chose Dice/kJ (kilojoules) as the metric describing this relationship. 

\section{Results and Discussion}

\subsection{Initial configuration and model architecture choice}

At first, a baseline run with Unet architecture was performed. Initial trials have shown that Attention-Squeeze-Unet performs almost on par with the baseline, while greatly shortening the training process. Consequently, all the following energy efficiency methods were evaluated with this model. The models were optimized using the Adam algorithm with a learning rate of 0.001. DDP in the default version was chosen for the communication between GPUs. Floating point operations were reduced to 16-bit precision. Until now, no other techniques have been applied. \\
We decided to use an early stopping algorithm to prevent overfitting and stop training when no improvement in validation loss is achieved in 15 consecutive epochs. The best parameters (that resulted in the lowest validation loss) of the model are saved and used later in the inference.\\
The batch size was set to 128 images per device (GPU). Training with a larger batch size would definitely be possible, however, using a too-large value can cause training instability and require tuning of other parameters that would offset the large gradient values effects.

\subsection{Analysis of Optimization Techniques and Results}

As the solutions were evaluated in an HPC cluster environment with significant available compute resources, data caching was the first efficiency technique explored. Using Monai CacheDataset abstraction, caching training and validation data led to shortening the time of epochs by nearly 400\% - a significant improvement that offsets the initial cost of loading cache within a few epochs. \\
The use of Novograd optimizer versus the Adam originally used has shown improvements in training speed. \\
We also experimented with setting different parameters $\beta_1$ and $\beta_2$. This led to slower training, drastically increasing the energy cost. These parameters could probably be tuned via an extensive hyperparameter search; however, energy constraints do not allow this in this setting. \\
The learning rate was chosen to be configured by automatic search. We decided to incorporate this step despite the benefits of using Novograd optimizer, which led to 2- 3\% improvements in final performance. However, increasing the search time above 100 training steps did not improve performance while increasing energy consumption. We also decided to incorporate learning rate scheduling and performing the learning rate decay by a factor of 10 when no improvement in validation loss was observed for the past 5 epochs, up to a minimum value of $10^{-6}$. This also led to a slight improvement in final performance, increasing the test Dice score by another 1-2\%. \\
The next step was to choose the number of workers to load the data. Even if the data is already in RAM and only transformations have to be applied, increasing the number of workers may still prove beneficial. Usually, the code would need to be profiled to determine the optimal number of workers, however, in the energy-constrained environment, it is impossible. We therefore decided to check 4, 6, and 8 workers for data loader. These experiments were done with every worker pre-fetching 2 batches of data. Six workers provided the best performance to energy consumed ratio. Going beyond a certain number of workers seems to decrease performance due to communication overhead, even if a suitable number of processor cores are available. Finally, we also decided to use pinned memory and persistent workers, as the memory resources allow for. This can also increase the throughput by removing the need to spawn worker processes every epoch and allocating memory for CPU to GPU data transfer.

Next, we add augmentation of the training images via rotation and random flipping. It led to improved performance at virtually no cost, increasing the test Dice score by nearly 5\%. 
Up to this point, the experiments were conducted using a Dice loss. We evaluated the usefulness of MCC and BCE losses. They led to worse performance without any reduction in energy usage. \\
To further improve communication with DDP, we disabled the search for unused parameters with every training step and set the computational graph to static. These changes allowed to increase the model's throughput and reduce training time. Using a bucket view for gradient reduction between devices led to decreased performance without energy benefits.\\
For comparison with DDP, experiments were conducted with algorithms offered by Bagua. The speedup in our case was not significant. Compared to a properly configured basic DDP strategy, it has shown slightly lower performance and higher energy consumption. 
Next, we evaluated the usage of PowerSGD. We have observed significant performance drops, as the compression is lossy and its parameters need to be properly tuned to avoid these drops while maintaining the benefits of faster communication. 
Similar results were observed when we evaluated the SWA. Research has shown the benefits of increased performance with no performance loss; however, once again the length of the training had to be known to properly tune the algorithm's hyperparameters. We did not observe any performance benefits in our tests.\\
We also examined the effects of pruning. Iterative pruning of the least important parameters evaluated using DeGraph was performed. During tests, we observed drastic drops in model performance. We therefore concluded that pruning on models that are already relatively small (less than 10M parameters) has to be done with caution, as the capacity of the model to learn the patterns in the data can turn out to be too low after applying the technique. 

\subsection{Final model choice}
\begin{figure}
    \centering
    \makebox[\textwidth]{\includegraphics[width=0.6\paperwidth]{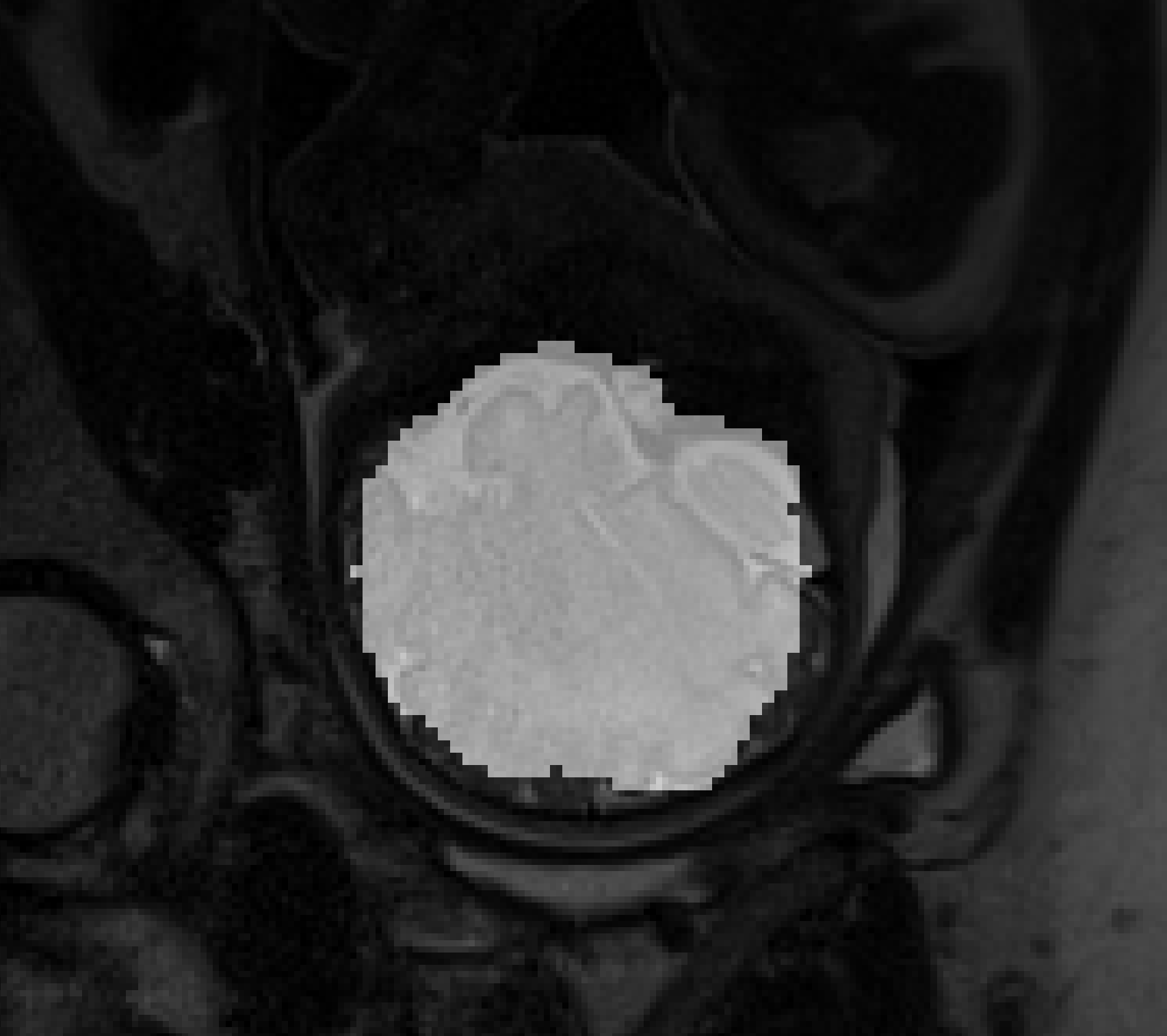}}
    \caption{Overlay between an original MRI slice and a segmented fetal  brain.}
    \label{fig:screenshot}
\end{figure}

\begin{figure}
    \centering
    \makebox[\textwidth]{\includegraphics[width=0.95\paperwidth]{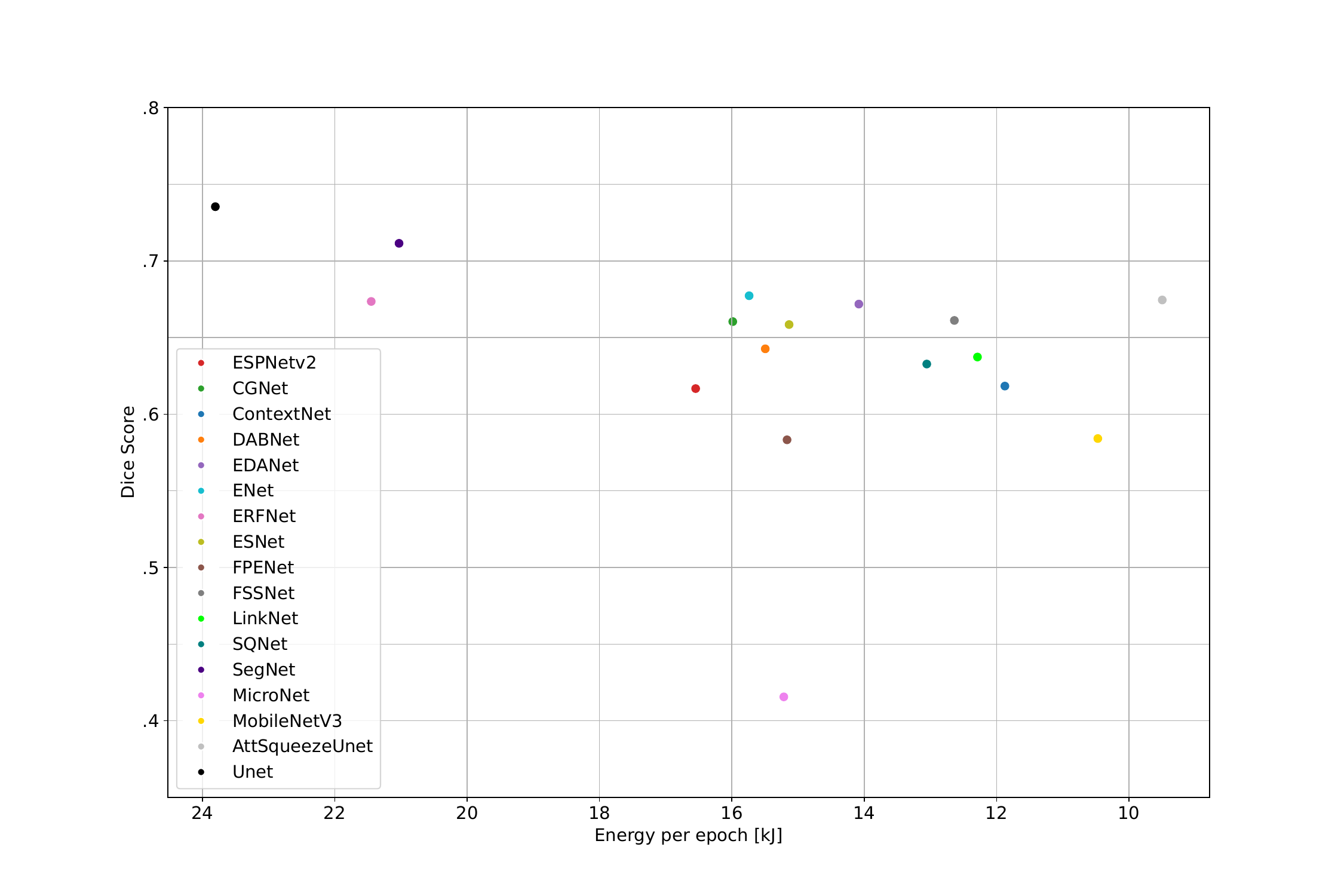}}
    \caption{Relationship of test Dice score in relation to the energy consumed per epoch of training.}
    \label{fig:Dice-energy-epoch}
\end{figure}

\begin{figure}
    \centering
    \makebox[\textwidth]{\includegraphics[width=0.95\paperwidth]{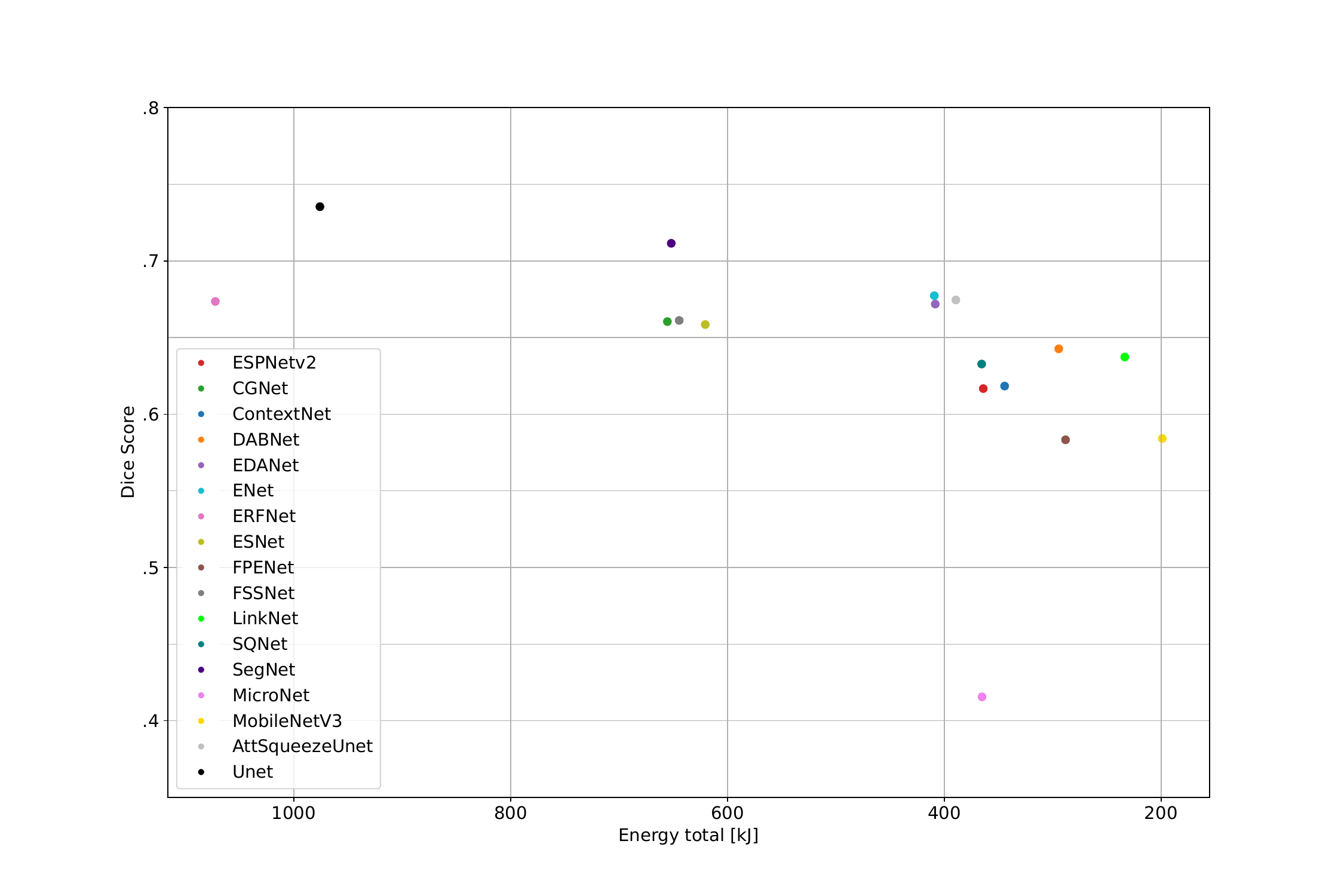}}
    \caption{Relationship of test Dice score in relation to the energy consumed during training.}
    \label{fig:Dice-energy-total}
\end{figure}

\begin{figure}
    \centering
    \makebox[\textwidth]{\includegraphics[width=0.7\paperwidth]{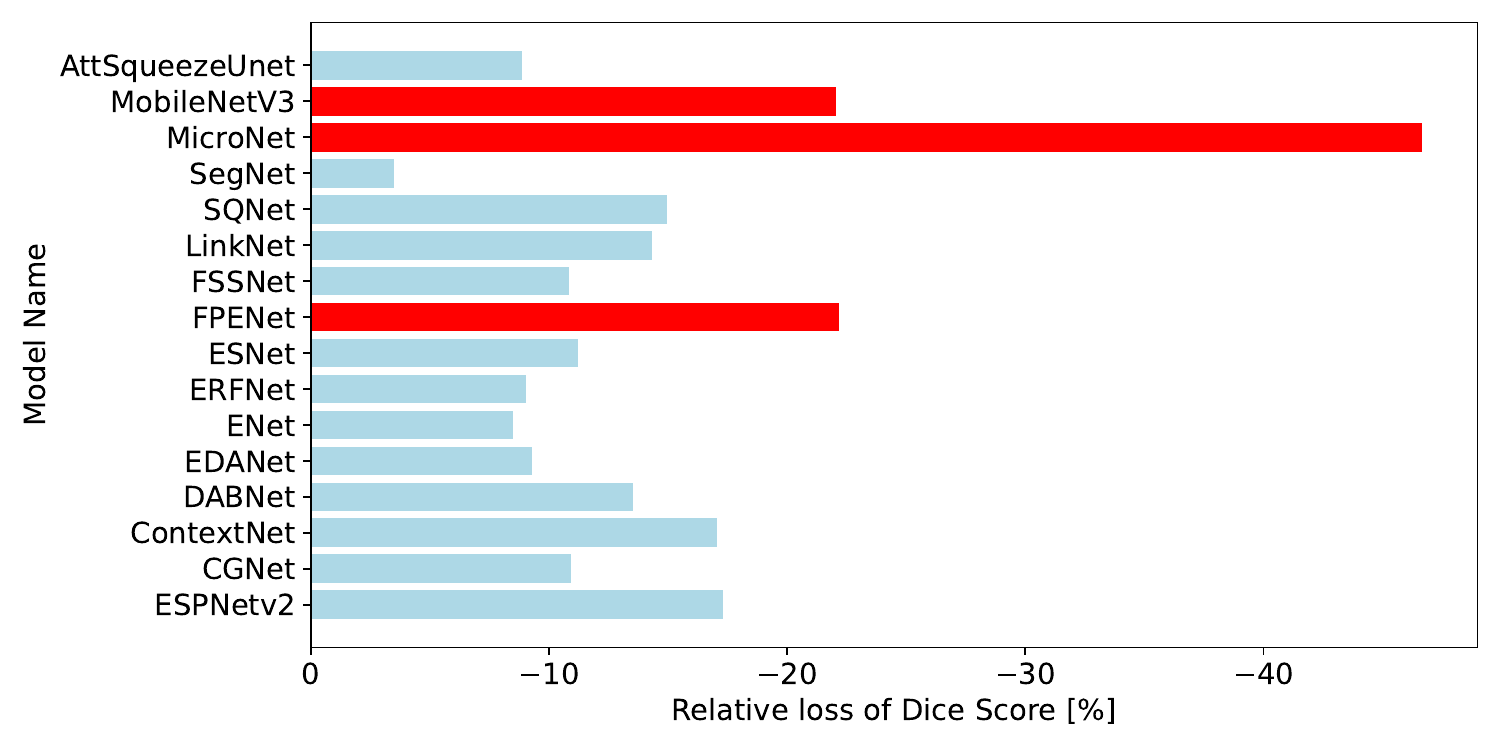}}
    \caption{Percentage of lost test Dice score for a given model tested relative to baseline value obtained by Unet.}
    \label{fig:Dice-relative-loss}
\end{figure}

As the pipeline and methods used were configured, we evaluated all the models listed. On the basis of these findings, we chose Attention-Squeeze-Unet as the final. The model has reached a high Dice score on the test data set relative to the baseline Unet. It has also shown the best energy per epoch ratio. To justify our choice, we present graphs of the performance of the model relative to energy consumption in Figures \ref{fig:Dice-energy-epoch} to \ref{fig:Dice-relative-loss}. We also report the segmentation to highlight qualitatively how noisy were the data, as depicted in Figure \ref{fig:screenshot}, where the overlay between an original MRI slice and a segmented results are showing, pointing out how cumbersome the data were and how challenging would be to obtain higher Dice score.

\subsection{Conclusions and recommendations}
We have shown that proper use of available methods can lead to satisfactory model performance while maintaining low energy consumption when training DNNs for medical image segmentation. In this study, we focused mostly on the training part of the process, since inference uses negligible amounts of energy when done only once. If the model is to be deployed and perform inference many times, then methods that were deemed too costly to optimize, such as pruning and quantization, can be considered, as the gains made for long-term operations can be substantial.

Recommendations for Energy-Efficient Fetal Brain Segmentation:
\begin{itemize}
    \item Optimization of data loading - configuration of the data loading pipeline should be considered every time. Caching provides the most significant speedups but should be used with caution only in environments with no memory constraints.
    \item Choice of optimizer - modern optimizers reach the performance on par with the established ones, while allowing for memory footprint reduction. Adaptive methods also seem to be robust to the hyperparameter choice, leading to a lower energy used for their configuration.
    \item Optimal distributed strategy and reducing floating point operation precision can offer significant throughput improvements without loss in performance. 
    \item Model architecture - using already existing or creating custom architecture that uses small amount of parameters leads to faster training, smaller compute requirements, and potentially still maintains satisfying levels of performance.
    \item Usage of methods requiring parameter tuning should be considered when compute resources allow for their tuning, otherwise they may result in suboptimal performance.
\end{itemize}

In summary, by enhancing the efficiency of machine learning algorithms through optimization techniques,  the computational demands can be significantly reduced. This not only can accelerate model training, but also minimizes the carbon footprint associated with the vast computational resources required. Taking into consideration this goal, we investigated and recommended some optimization ideas to reduce energy consumption and carbon emissions. We would like to stress the importance of sustainable computational approaches, and we would like to invite the scientific community to focus further on care about carbon footprints.

\begin{credits}
\subsubsection{\ackname} This publication is supported by the European Union’s Horizon 2020 research and innovation programme under grant agreement Sano No 857533.
This publication is supported by Sano project carried out within the International Research Agendas programme of the Foundation for Polish Science, co-financed by the European Union under the European Regional Development Fund. 
This research was supported in part by the PLGrid infrastructure on the Athena computer cluster.

\end{credits}
%
%
%
\bibliographystyle{splncs04}
\bibliography{references}

\end{document}